# Stabilization of layered perovskite structures via strontium substitution in $Ca_3Ti_2O_7$ revealed via elemental mapping

Kosuke Kurushima[1], Hiroshi Nakajima[2], Shinya Mine[3], Hirofumi Tsukasaki[2], Masaya Matsuoka[4], Bin Gao[5, 6], Sang-Wook Cheong[5], and Shigeo Mori[2, *]

[1]*Toray Research Center, Ohtsu, Shiga 520-8567, Japan.*

[2]*Department of Materials Science, Osaka Prefecture University, Sakai, Osaka 599-8531, Japan.*

[3]*Institute for Catalysis, Hokkaido University, N-21, W-10, Sapporo 001-0021, Japan.*

[4]*Department of Applied Chemistry, Osaka Prefecture University, Sakai, Osaka, 599-8531, Japan.*

[5]*Rutgers Center of Emergent Materials, Rutgers University, Piscataway, New Jersey 08854, USA.*

[6]*Present address: Department of Physics and Astronomy, Rice University, Houston, Texas 77005-1827, USA.*

*e-mail: mori@mtr.osakafu-u.ac.jp*

Extensive studies have been performed on layered compounds, ranging from layered cuprates to van der Waals materials with critical issues of intergrowths and stacking faults. However, such structures have been studied less because of experimental difficulty. We present characteristic defect structures of intergrowths in the Ruddlesden–Popper $Ca_{2.46}Sr_{0.54}Ti_2O_7$, which is known to exhibit hybrid improper ferroelectricity. Transmission electron microscopy reveals that numerous intergrowths composed of seven and 15 layers are introduced in the ferroelectric domains. Elemental maps demonstrate that Sr ions are selectively substituted into the perovskite layers of intergrowths. Density functional theory calculations support the site-selective substitution of Sr ions, favorably located in the intergrowths. The stabilization of the Ruddlesden–Popper phase and intergrowths via Sr substitution can be explained by the ionic-radius difference between Ca and Sr ions. The study reveals detailed defect structures originating from the layered perovskite structure of $Ca_{2.46}Sr_{0.54}Ti_2O_7$, and shows the usefulness of elemental mapping in probing the substitution effects in oxides.

## Introduction

Ferroelectrics are used in science and medical fields and several other industrial applications. $BaTiO_3$ and $Pb(Zr_xTi_{1-x})O_3$ are typical ferroelectric materials used as commercial capacitors and filters. For example, $BaTiO_3$ is the main material in multilayer ceramic capacitors, and $Pb(Zr_xTi_{1-x})O_3$ is used in devices utilizing elastic vibration, such as actuators. Recently, many ferroelectric materials have been predicted based on theoretical calculations, and various ferroelectric materials are found in the perovskite ($ABO_3$) and layered perovskite structures[1–5].

$Ca_3Ti_2O_7$ is predicted to exhibit ferroelectricity as per the first-principles calculations[6]. According to the previous studies, $Ca_3Ti_2O_7$ and $Ca_3Mn_2O_7$ ($A_{n+1}B_nO_{3n+1}$; $n = 2$) compounds exhibit the Ruddlesden–Popper structure and structural phase transition from the paraelectric phase with space group *I4/mmm* to the ferroelectric phase with space group *A2₁am*. The materials exhibit spontaneous ferroelectric polarization along the [100] axis[7–9]. As shown in Figure 1, the orthorhombic crystal with space group *A2₁am* (lattice constants: $a = 5.4234$ Å, $b = 5.4172$ Å, $c = 19.5169$ Å, $\alpha = \beta = \gamma = 90°$) consists of two kinds of layers, a perovskite block layer with $TiO_6$ octahedron and a rock-salt block layer shared at the vertex. The characteristic crystal structure leads to ferrielectricity and polarization parallel to [100] axis due to $TiO_6$ octahedron rotation and the inclination around the [100] axis[10]. The polarization mechanism is called hybrid improper ferroelectricity[6,11,12].

$Ca_{2.46}Sr_{0.54}Ti_2O_7$ exhibits an interesting ferroelectric domain structure composed of various head-to-head and tail-to-tail type charged domain walls[10]. In $Ca_3Ti_2O_7$, eight types of ferroelectric domains are formed by a combination of rotation and gradient directions of $TiO_6$ octahedron, and four kinds of ferroelectric domains are permitted within one ferroelastic domain[13]. The charged domain wall is observed only in limited cases because of the large energy cost[14,15]. Recently, high-angle annular dark-field scanning transmission electron microscopy (HAADF-STEM) was conducted and the details of the charged domain wall structure were reported[16–18]. Another study demonstrated that the 180° domain walls and charged domain walls are influenced by intergrowths[19]. The intergrowths are the defects occurring in the layered perovskite structure as the irregular stacking of layers. Besides, these intergrowths and stacking faults have been sought in many layered materials, such as cuprates and van der Waals crystals as they affect the physical properties[20–22]. However, direct imaging, including elemental mapping is challenging for atomic-scale defects. In layered perovskite structures, the detailed intergrowths have not been revealed although their presence has been known for decades[23,24].

In this study, defect structures of intergrowths were examined in the Sr-substituted $Ca_3Ti_2O_7$ ($Ca_{2.46}Sr_{0.54}Ti_2O_7$) via the atomic-resolution HAADF-STEM combined with energy-dispersive x-ray spectroscopy (EDS). In addition to characteristic ferroelectric domain structures, site-selective substitution of intergrowths was found in improper ferroelectric $Ca_{2.46}Sr_{0.54}Ti_2O_7$.

## Methods

High-resolution HAADF-STEM with EDS revealed the microstructures of $Ca_{2.46}Sr_{0.54}Ti_2O_7$. The atomic-resolution EDS mapping can distinguish elements, that are difficult to detect via HAADF-STEM[25–27]. A JEOL JEM-ARM200F STEM with a spherical-aberration corrector and a dual EDS system with double 100 mm$^2$ silicon drift detectors was used at the acceleration voltage of 200 kV. Atomic-resolution EDS maps were acquired via the NSS3.3 software (ThermoFisher Scientific Inc.). EDS spectral images were acquired with an electron probe of 1.2 Å, convergence semi-angle of the incident electron beam of 22 mrad, and current of 60 pA. HAADF-STEM images were recorded under the similar optical condition using an annular detector with a semi-angle of the detector, ranging from 90 to 170 mrad. Spectral images were acquired as a series of frames, where the same region was scanned multiple times. The total acquisition time was about 16 min with a dwell time of 20 μs and a typical frame of 256 × 256 pixels. The obtained EDS images were subjected to the Fourier filtering process for noise reduction. STEM observation and density functional theory (DFT) calculation methods in the present study are similar to those reported in Ref[18]. All measurements were done at room temperature. $Ca_{2.46}Sr_{0.54}Ti_2O_7$ single crystal sample was prepared by the floating zone method similar to a previous study[10]. Briefly, CaO, SrO, and $TiO_2$ were weighed and ground with a mortar. The powdered samples were pressed into a rod by applying hydrostatic pressure. The rod was sintered at 1200°C for 15 h. Partial area of the rod was melted with a halogen-lamp furnace and the molten zone was moved at approximately 3 mm/h to grow the single crystal. The sample for transmission electron microscope (TEM) observation was prepared by focused ion beam (FIB), which initially produced a TEM specimen of thickness less than 100 nm. Subsequent Ar-ion milling was performed to remove the FIB damage on the specimen.

**Results and discussion**

Figure 2a shows the dark-field image of $Ca_{2.46}Sr_{0.54}Ti_2O_7$ along the [010] axis, captured using reflection with a scattering vector of $\vec{g}$ = 202. By using the dynamic effect producing the failure of Friedel's law, the dark-field image was obtained under the two-beam condition, wherein a ferroelectric domain with $\vec{g} \cdot \vec{P} > 0$ produces bright areas in the image[28,29] ($\vec{g}$ and $\vec{P}$ are the scattering and polarization vectors, respectively). Thus, bright and dark areas correspond to ferroelectric domains with the right [100] and left [$\bar{1}$00] polarization directions, respectively. The image depicts wavy charged domain walls and straight defects marked by yellow lines. The straight defects are intergrowths, as explained later. Figure 2b–f shows the HAADF-STEM and EDS images in the right polarized domain of Figure 2a. Figure 2b shows five layers of ions, wherein small displacements of ions are shown by the arrows. The EDS images demonstrate that the five layers comprise a sequence of Ca, Ti, Sr, Ti, and Ca ions. The dark layers in the HAADF-STEM image correspond to the rock-salt layers among the Ca (green) ions. Moreover, Ca ions tend to occupy the rock-salt sites, while almost all Sr ions are located in perovskite sites. Ti ions are located in the oxygen octahedral sites of the perovskite layer. The result agrees with the previous study of electron energy-loss spectroscopy[16].

Considering the composition of $Ca_{2.46}Sr_{0.54}Ti_2O_7$, the ratio of Ca and Sr is approximately 5:1. Conversely, the ratio of sites for the alkaline-earth-metal elements (Ca and Sr) in the rock-salt and perovskite layers is 2:1. Thus, half of the perovskite layer sites are occupied by Ca ions with Sr ions occupying the other half. Other Ca ions occupy the rock-salt layer sites if Sr ions preferentially occupy the perovskite layer sites. Consistent with this prediction, the perovskite layer sites show both Ca and Sr ions. Previous DFT studies show that Sr ions preferably occupy the perovskite layer sites[18,30]. The observation results demonstrate that this site selection occurs in the Sr-substituted $Ca_3Ti_2O_7$.

Next, the region around the defect in a charged domain wall marked in yellow was investigated. The observation is reflected in Figure 3a, which can be distinguished by antipolar displacements[16]. The area includes the $n = 3$ (seven layers) intergrowth structure (Figure 3b). Note that Ruddlesden–Popper oxides can be expressed as $A_nB_nO_{3n}$–$AO$ (where $A$ is an alkaline-earth-metal ion, $B$ is a transition metal ion, and $n$ is the number of simple perovskite-type unit cells), and thus, the intergrowth in Figure 3b corresponds to the $n = 3$ structure[19]. Such intergrowths are sometimes observed in layered perovskites[23,24]. Figure 3c–f shows the corresponding atomic-resolution elemental maps. The images reveal that the intergrowth comprises three perovskite blocks of Sr and Ti ions without the rock-salt layer. Hence, the site selection for the perovskite layer is also valid for the $n = 3$ intergrowth.

DFT calculations were performed to evaluate the Sr stability in the perovskite and rock-salt sites in the $n = 3$ intergrowth. Supplementary Figure 1 illustrates the model of the Ruddlesden–Popper $n = 3$ intergrowth structure with the Sr ions located in the perovskite or rock-salt sites. The model is based on the structural analysis[31]. The total energy was calculated after structural relaxation using the following equation:

$$\Delta H = \frac{1}{N}\{E_{CSTO} + E_{Ca} - E_{Sr} - E_{CTO}\} \quad .(1)$$

Here, $N = 68$ is the total atoms in the unit cell. The first, second, third, and fourth terms represent the energies of Sr-substituted $Ca_3Ti_2O_7$, face-centered Ca, face-centered Sr, and $Ca_3Ti_2O_7$. As listed in Supplementary Table 1, the calculations show that the structure substituted by Sr in A site of the perovskite layer has 2.77 eV/atom, whereas A' site of the rock-salt layer has 5.11 eV. Like the $n = 2$ structure, the perovskite site is more stable than the rock-salt site by 2.34 eV/atom for Sr ions in the $n = 3$ Ruddlesden–Popper structure. DFT calculations support the result that the perovskite site is more stable than the rock-salt site for Sr ions in the $n = 3$ structure.

Furthermore, high-resolution HAADF and EDS images were obtained to confirm the tendency of site occupancy in higher-number intergrowth. Figure 4 unambiguously delineates the site selection of Sr ions in the $n = 7$ intergrowths because it has 15 layers in the perovskite layers. Similar to the results above, the intergrowth comprises a series of Sr and Ti ions in the perovskite layer. These results reflect the site-selective substitution for the high-number intergrowth.

The observation results show that the intergrowths are formed by Sr substitution as Ca ions can occupy the sites of the rock-salt layer and Sr ions prefer to occupy the sites of the perovskite layer. The intergrowth structure is a simple perovskite structure without the rock-salt layer. DFT calculations also show that Sr ions prefer the perovskite layer sites. Thus, Sr substitution should induce the intergrowths, as observed in this study. The tendency is also validated by the results that the intergrowths of any numbers are composed of Sr-rich layers. Notably, the intergrowths could not change macroscopic properties, such as the dielectric constant and transition temperature due to small volume fraction of intergrowths. However, the intergrowths may affect the local ferroelectric phenomena. As shown in Figure 2a, 180° domains are separated by intergrowths. Thus, some ferroelectric domains with intergrowths can be narrower than those without intergrowth. In fact, a narrow ferroelectric domain of a black area below the mark $P_s$ is present in Figure 2a. Additionally, ferroelectric domain walls move along the intergrowth lines when an electric field is applied because some domain walls are terminated and pinned by the intergrowths, as shown in Figure 2a.

Empirically, growing a large single crystal of $Ca_3Ti_2O_7$ is difficult. However, Sr substitution makes it easier and stabilizes the $n = 2$ Ruddlesden–Popper phase, possibly due to the stabilization of the layered perovskite structures by Sr ions, while Ca ions are not stable in the sites of the perovskite layer, as demonstrated in this study. The ionic radius of $Sr^{2+}$ ($r_{sr}$ = 1.58Å) is larger than that of $Ca^{2+}$ ($r_{ca}$ = 1.48Å). The sites in perovskite layers are bigger than rock-salt layers, and Ca ions are small to occupy the perovskite sites. In other words, the tolerance factors $t = (r_A + r_O)/\sqrt{2} (r_B + r_O)$ in Sr and Ca are 1.00 and 0.966, respectively. ($r_A$, $r_B$, and $r_O$ represent the ionic radii of $A$ site, $B$-site ($Ti^{4+}$), and $O^{2-}$ ions in the perovskite $ABO_3$, respectively.) In fact, bulk $SrTiO_3$ has a cubic structure of $Pm\bar{3}m$ while bulk $CaTiO_3$ has an orthorhombic structure of $Pbnm$ with oxygen octahedral rotations.[32,33] The experimental evidence suggests that Sr ions tend to make the layered perovskite structure more stable than Ca. However, the difference in the ionic radii between $Sr^{2+}$ and $Ca^{2+}$ is not so large, indicating the small stability of $Sr^{2+}$ relative to $Ca^{2+}$ in the perovskite layers. Therefore, Sr and Ca ions should be frequently occupied at both perovskite and rock-salt layers: The two ions could be randomly replaced with each other when the specimen was melted during the single crystal growth. This was confirmed by the intensity profile shown in the supplementary Figures 2 and 3, wherein Sr and Ca ions were observed in the rock-salt and perovskite layers as small intensity distributions.

## Conclusions

To elucidate the effect of substituting Sr ions at the Ca site, HAADF-STEM combined with EDS was applied to $Ca_{2.46}Sr_{0.54}Ti_2O_7$ because high-resolution HAADF-STEM observation with EDS analysis was highly effective in understanding the local defect structures in ferroelectric materials. This study revealed that the layered perovskite structure of the Ruddlesden–Popper phase has numerous intergrowths in improper ferroelectric $Ca_{2.46}Sr_{0.54}Ti_2O_7$. The EDS maps demonstrate that the intergrowths are composed of Sr and Ti ions. The intergrowths are stabilized by Sr ions as they prefer the perovskite layer sites, as confirmed by DFT calculations. The large Sr ionic radius should contribute to the stabilization of the $n = 2$ Ruddlesden–Popper phase in crystal growth. The results indicate that the stabilization of the Ruddlesden–Popper structure is influenced by the substitution elements.

**References**

[1] C.J. Fennie and K.M. Rabe, Phys. Rev. Lett. **97**, 267602 (2006).
[2] C.J. Fennie, Phys. Rev. Lett. **100**, 167203 (2008).
[3] E. Bousquet, M. Dawber, N. Stucki, C. Lichtensteiger, P. Hermet, S. Gariglio, J.-M. Triscone, and P. Ghosez, Nature **452**, 732 (2008).
[4] A. Roy, J.W. Bennett, K.M. Rabe, and D. Vanderbilt, Phys. Rev. Lett. **109**, 37602 (2012).
[5] N.A. Pertsev, A.K. Tagantsev, and N. Setter, Phys. Rev. B **61**, R825 (2000).
[6] N.A. Benedek and C.J. Fennie, Phys. Rev. Lett. **106**, 107204 (2011).
[7] W. Zhu, L. Pi, Y. Huang, S. Tan, and Y. Zhang, Appl. Phys. Lett. **101**, 192407 (2012).
[8] M.A. Green, K. Prassides, P. Day, and D.A. Neumann, Int. J. Inorg. Mater. **2**, 35 (2000).
[9] M. V Lobanov, M. Greenblatt, N.C. El'ad, J.D. Jorgensen, D. V Sheptyakov, B.H. Toby, C.E. Botez, and P.W. Stephens, J. Phys. Condens. Matter **16**, 5339 (2004).
[10] Y.S. Oh, X. Luo, F.-T. Huang, Y. Wang, and S.-W. Cheong, Nat. Mater. **14**, 407 (2015).
[11] N.A. Benedek, A.T. Mulder, and C.J. Fennie, J. Solid State Chem. **195**, 11 (2012).
[12] A.T. Mulder, N.A. Benedek, J.M. Rondinelli, and C.J. Fennie, Adv. Funct. Mater. **23**, 4810 (2013).
[13] F.-T. Huang, F. Xue, B. Gao, L.H. Wang, X. Luo, W. Cai, X.-Z. Lu, J.M. Rondinelli, L.Q. Chen, and S.-W. Cheong, Nat. Commun. **7**, 11602 (2016).
[14] G. Catalan, J.F. Scott, A. Schilling, and J.M. Gregg, J. Phys. Condens. Matter **19**, 22201 (2006).
[15] R.C. Miller and G. Weinreich, Phys. Rev. **117**, 1460 (1960).
[16] M.H. Lee, C.-P. Chang, F.-T. Huang, G.Y. Guo, B. Gao, C.H. Chen, S.-W. Cheong, and M.-W. Chu, Phys. Rev. Lett. **119**, 157601 (2017).
[17] K. Kurushima, W. Yoshimoto, Y. Ishii, S.-W. Cheong, and S. Mori, Jpn. J. Appl. Phys. **56**, 10PB02 (2017).
[18] H. Nakajima, K. Kurushima, S. Mine, H. Tsukasaki, M. Matsuoka, B. Gao, S. Cheong, and S. Mori, Commun. Mater. **2**, 109 (2021).
[19] H. Nakajima, K. Shigematsu, Y. Horibe, S. Mori, and Y. Murakami, Mater. Trans. **60**, 2103 (2019).
[20] B. Domenges, M. Hervieu, C. Martin, D. Bourgault, C. Michel, and B. Raveau, Phase Transitions A Multinatl. J. **19**, 231 (1989).
[21] C.N.R. Rao, Mater. Sci. Eng. B **18**, 1 (1993).
[22] A. Lotnyk, T. Dankwort, I. Hilmi, L. Kienle, and B. Rauschenbach, Scr. Mater. **166**, 154 (2019).
[23] J. Sloan, P.D. Battle, M.A. Green, M.J. Rosseinsky, and J.F. Vente, J. Solid State Chem. **138**, 135 (1998).
[24] P.D. Battle and M.J. Rosseinsky, Curr. Opin. Solid State Mater. Sci. **4**, 163 (1999).
[25] M.-W. Chu, S.C. Liou, C.-P. Chang, F.-S. Choa, and C.H. Chen, Phys. Rev. Lett. **104**, 196101 (2010).
[26] A.J. d'Alfonso, B. Freitag, D. Klenov, and L.J. Allen, Phys. Rev. B **81**, 100101 (2010).
[27] P. Lu, J. Xiong, M. Van Benthem, and Q. Jia, Appl. Phys. Lett. **102**, 173111 (2013).
[28] M. Tanaka and G. Honjo, J. Phys. Soc. Japan **19**, 954 (1964).
[29] F. Fujimoto, J. Phys. Soc. Japan **14**, 1558 (1959).
[30] C.F. Li, S.H. Zheng, H.W. Wang, J.J. Gong, X. Li, Y. Zhang, K.L. Yang, L. Lin, Z.B. Yan, S. Dong, and others, Phys. Rev. B **97**, 184105 (2018).
[31] P.D. Battle, M.A. Green, J. Lago, J.E. Millburn, M.J. Rosseinsky, and J.F. Vente, Chem. Mater. **10**, 658 (1998).
[32] W.-Q. Luo, Z.-Y. Shen, Y.-M. Li, Z.-M. Wang, R.-H. Liao, and X.-Y. Gu, J. Electroceramics **31**, 117 (2013).
[33] M. Yashima and R. Ali, Solid State Ionics **180**, 120 (2009).

**Figures**

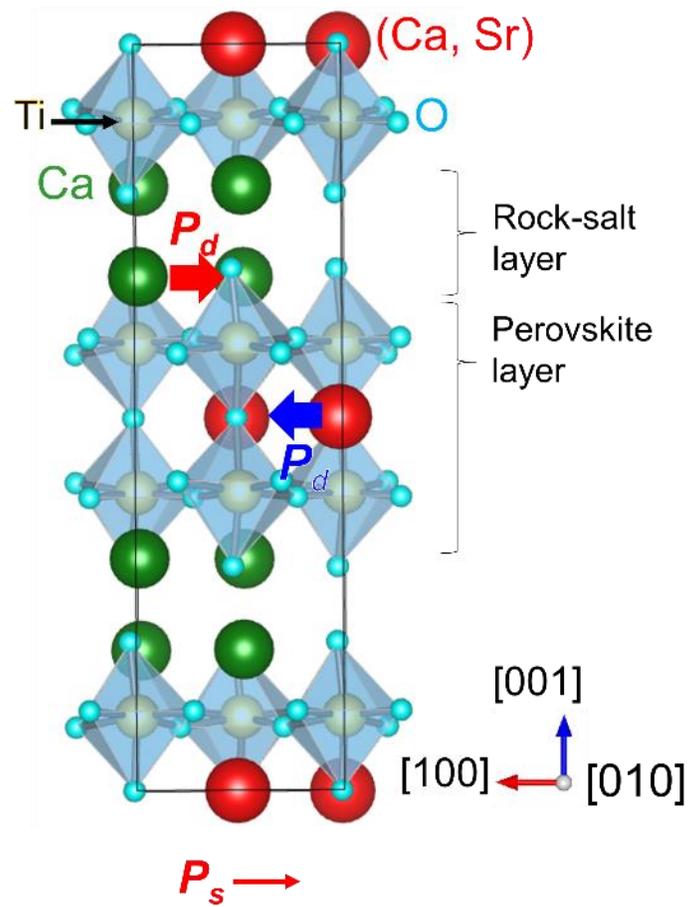

Fig. 1. Crystal structure along the [010] axis. This direction corresponds to the observation condition in this study. Note that occupancy of both Ca and Sr in the perovskite site are 0.5 in the $Ca_{2.5}Sr_{0.5}Ti_2O_7$ composition if all rock-salt sites are occupied by Ca ions and since rock-salt and perovskite layers have eight and four sites in the unit cell, respectively.

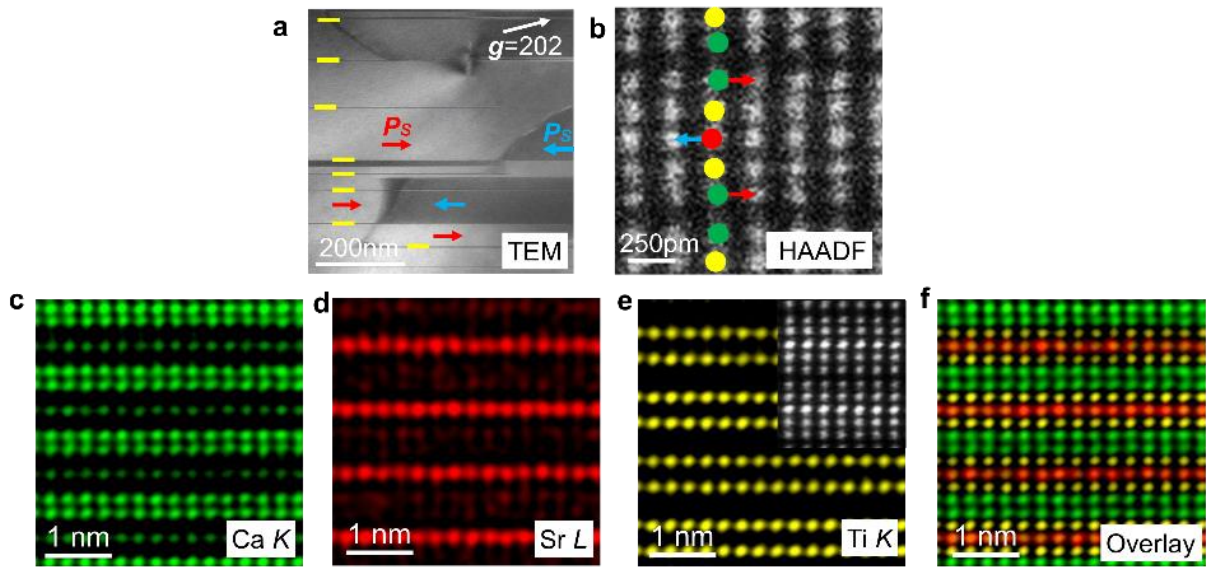

Fig. 2. Atomic-resolution observation of $Ca_{2.46}Sr_{0.54}Ti_2O_7$. (a) Dark-field transmission electron microscopy (TEM) image of the (010) plane. Red and blue arrows represent the [100] and [$\bar{1}$00] polarization directions, respectively. Yellow bars show positions of intergrowths. (b) Atomic-resolution HAADF-STEM image in the right polarized domain-marked $Ps$ in (a). Green, red, and yellow spheres represent Ca, Sr, and Ti ions, respectively, and arrows show displacement directions. (c)–(e) EDS-STEM images of each element. Inset of (e) shows a corresponding HAADF-STEM image. (f) Overlay images of (c), (d), and (e). Labels $K$ and $L$ are energy levels used for mapping.

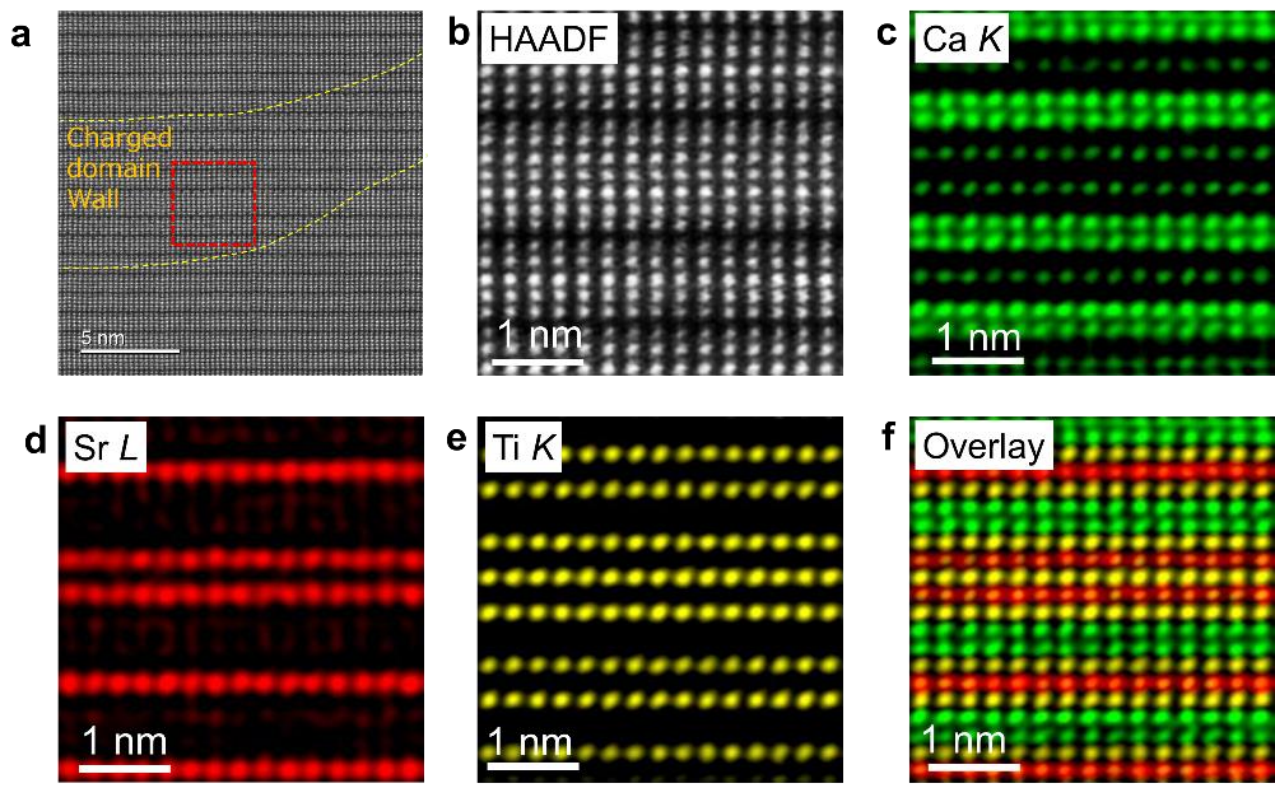

Fig. 3. Intergrowth structure in $Ca_{2.46}Sr_{0.54}Ti_2O_7$. (a) HAADF-STEM image in a charged domain wall. (b) High-magnification HAADF-STEM image in the marked area in (a), which depicts an intergrowth structure of $n = 3$ (seven atomic layers). (c)–(f) EDS images in the corresponding area in (b). Energy levels used for EDS images are displayed after the elements.

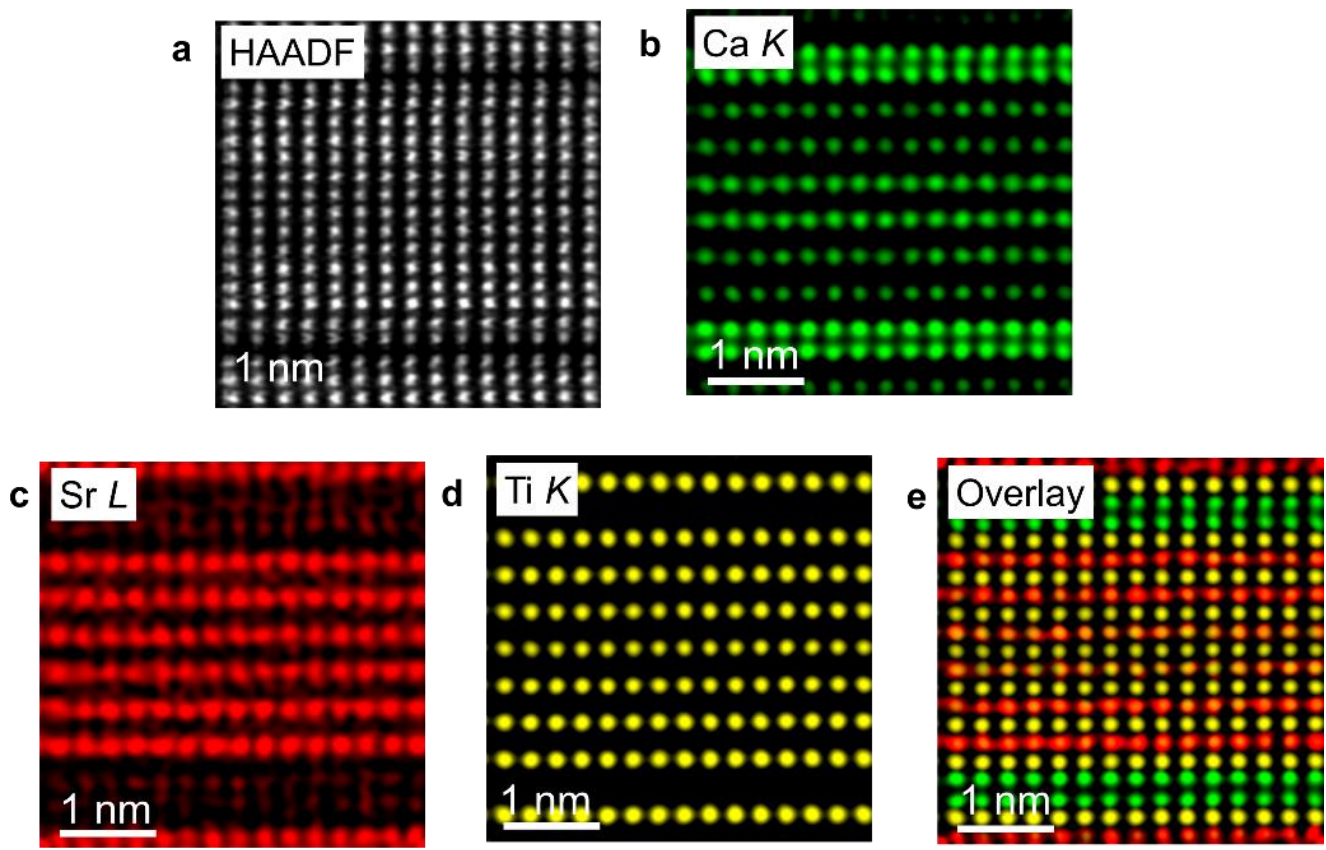

Fig. 4. Microstructure of an *n* = 7 intergrowth. (a) HAADF-STEM image around an *n* = 7 intergrowth (15 atomic layers). (b)–(e) EDS images corresponding to the HAADF-STEM image (a).

# Density functional theory calculation in an *n* = 3 intergrowth structure

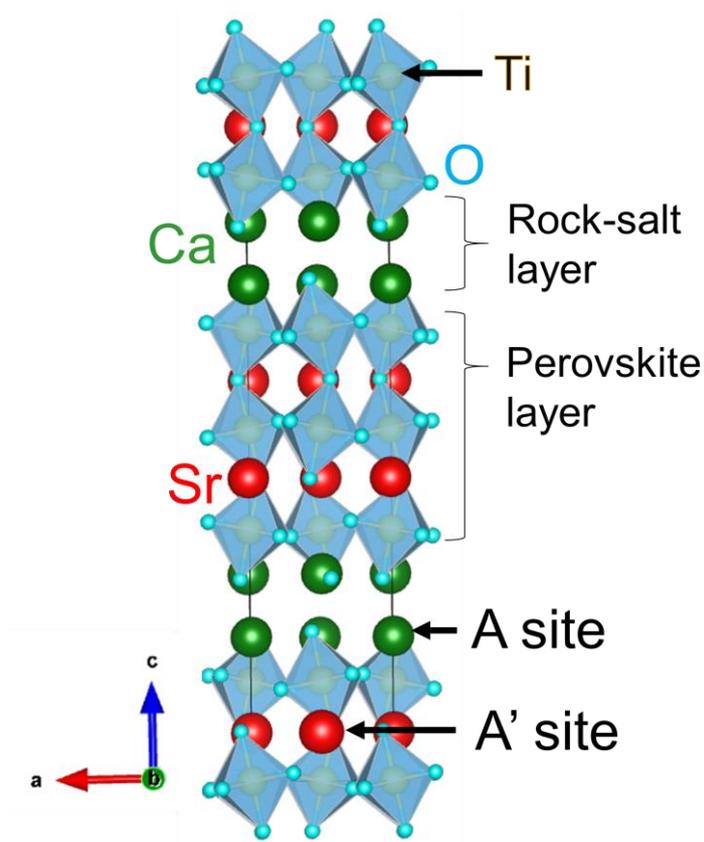

**Supplementary Fig. 1.** Structural model for calculating the substitution energy for the intergrowth structure of *n* = 3. Sr atom is substituted for Ca atom in the perovskite (A' site) and rock-salt (A site) layers.

**Supplementary Table 1.** Total energy of Sr substitution using density functional theory. Two cases were modeled: Sr ion is substituted in A (rock-salt) or A' (perovskite) sites, and the energies were calculated after structural relaxation. The term $E_{CTO}$ represents the energy of the structure-free Sr substitution. $E_{Ca}$ and $E_{Sr}$ show the energies of a single atom in the stable face-centered cubic structures.

|         | $E_{CSTO}$ | $E_{Ca}$ | $E_{Sr}$ | $E_{CTO}$ | $\Delta H$ (meV/atom) |
|---------|------------|----------|----------|-----------|------------------------|
| A site  | −511.107   | −1.92981 | −1.63803 | −511.746  | 5.10928                |
| A' site | −511.266   | −1.92981 | −1.63803 | −511.746  | 2.76575                |

**Intensity profile in the Ruddlesden-Popper structure**

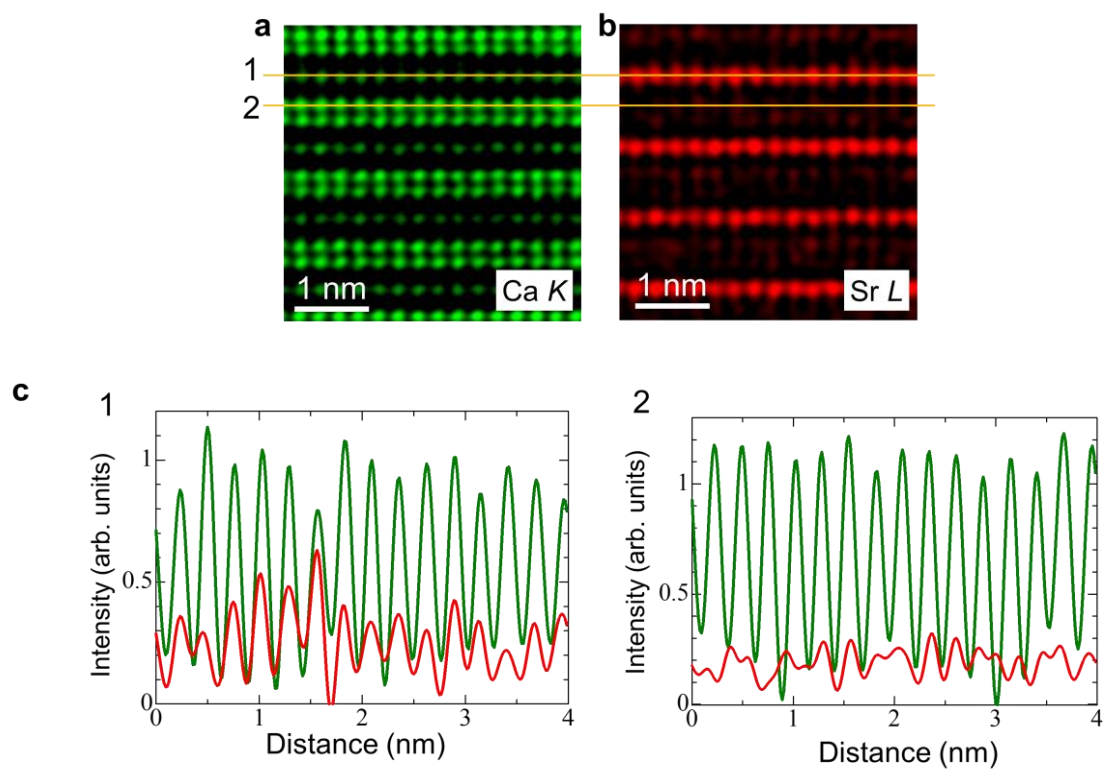

**Supplementary Fig. 2.** Intensity profile in the Ruddlesden-Popper structure. (a) and (b) Elemental maps using the Ca $K$ and Sr $L$ edges, respectively. These images are the same as those of Fig. 2. (c) Intensity profile along the lines 1 and 2. The lines 1 and 2 corresponds to the perovskite and rock-salt layers, respectively. The intensity profile shows that Sr ions occupy the sites of the rock-salt layers although the intensity is low. Note that the intensity of Ca is higher than that of Sr in the perovskite layer because the $K$ edge is used for imaging, although the ratio of Sr is higher in the perovskite layer.

**Intensity profile in the *n* = 3 intergrowth structure**

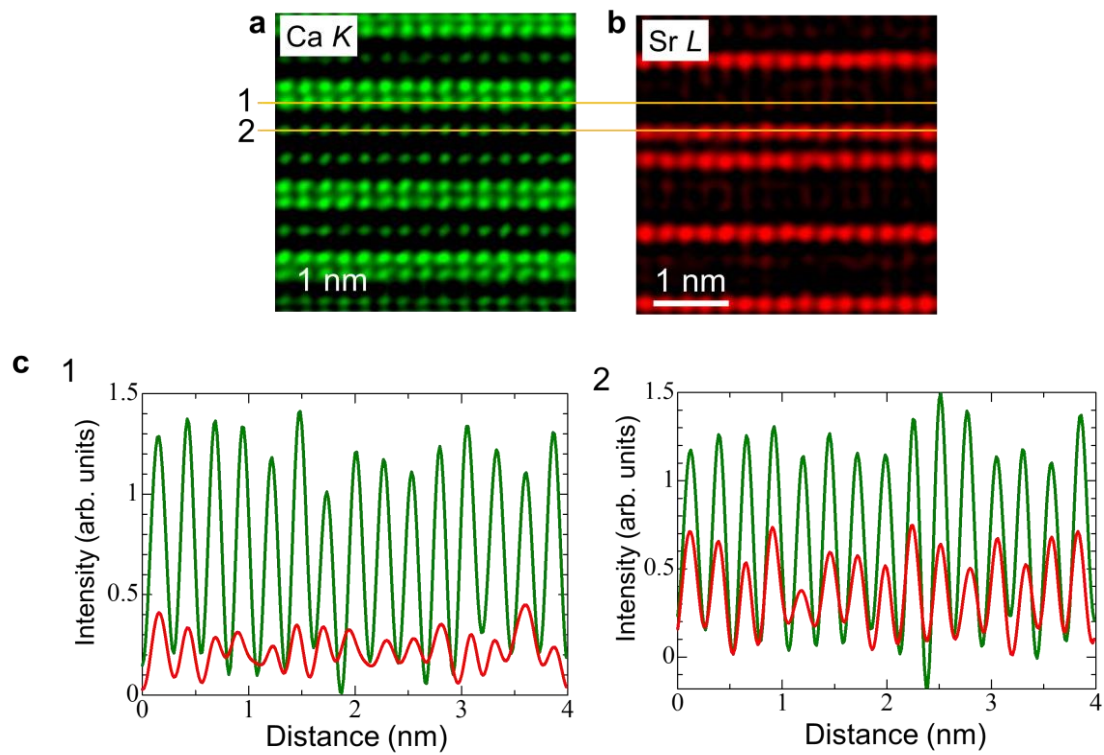

**Supplementary Fig. 3.** Intensity profile in the *n* = 3 intergrowth. (a) and (b) Elemental maps using the Ca *K* and Sr *L* edges, respectively. These images are the same as those of Fig. 3. (c) Intensity profile along the lines 1 and 2. The lines 1 and 2 corresponds to the perovskite and rock-salt layers, respectively. The intensity profiles show that Ca ions occupy the sites of the perovskite layers in the intergrowth.